\documentclass[aps,preprint,amssymb,12pt,floatfix]{revtex4}
\usepackage{subfigure}
\usepackage{graphicx}
\usepackage{lscape,graphicx}
\usepackage{rotating}
\usepackage{epstopdf}
\usepackage{color}
\usepackage{amsmath,amsthm}

\begin{document} 
\title{Helices 2 and 3 are the initiation sites in the  PrP$^C$ $\rightarrow$ PrP$^{SC}$ transition}
\author{Jie Chen$^1$ and D. Thirumalai$^{1,2}$}\thanks{Corresponding author phone: 301-405-4803; fax: 301-314-9404; thirum@umd.edu}
\affiliation{$^1$Biophysics Program,Institute for Physical Science and Technology
$^2$Department of Chemistry and Biochemistry, University of Maryland, College Park, MD 20742\\}

\date{\small \today}
 
\begin{abstract}
It is  established that prion protein is the sole causative agent in a number of diseases in humans and animals. However, the nature of conformational changes that the normal cellular form PrP$^C$ undergoes in the conversion process to a self-replicating state is still not fully understood. The  ordered C-terminus of  PrP$^C$ proteins has three helices (H1, H2, and H3). Here, we use the Statistical Coupling Analysis (SCA) to infer co-variations at various locations using a family of evolutionarily related sequences, and the response  of mouse and human PrP$^C$s to mechanical force to decipher the initiation sites for transition from PrP$^C$ to an aggregation prone PrP* state.   The sequence-based SCA predicts that the clustered residues in non-mammals are localized in the stable core (near H1) of PrP$^C$ whereas in mammalian PrP$^C$ they are localized in the frustrated helices H2 and H3 where most of the pathogenic mutations are found.   Force-extension curves and free energy profiles as a function of extension of mouse and human PrP$^C$ in the absence of disulfide (SS) bond between residues Cys179 and Cys214, generated by applying mechanical force to the ends of the molecule, show a sequence of unfolding events starting first with rupture of H2 and H3.  This is followed by disruption of structure in two strands. Helix H1, stabilized by three salt-bridges, resists substantial force before unfolding. Force extension profiles and the dynamics of rupture of tertiary contacts also show that even in the presence of SS bond the instabilities in most of H3 and parts of H2 still determine the propensity to form the PrP* state. In mouse PrP$^C$ with SS bond there are about ten residues that retain their order even at high forces.  Both SCA and single molecule force simulations show that  in the conversion process from PrP$^C$ to PrP$^{Sc}$ major conformational changes occur (at least initially) in H2 and H3, which due their sequence compositions are frustrated in the helical state.  Implications of our findings for structural model for the scrapie form of PrP$^C$ are discussed.
\end{abstract}

\maketitle

\section{Introduction}

Aggregation of misfolded proteins is implicated in a number of diseases\cite{Prusiner98PNAS,Aguzzi10NatRevDD}. For example, misfolding of the extracellular globular prion proteins, attached
to the plasma membrane by a glycosylphosphatidylinositol
anchor, is associated with a variety of  transmissible spongiform
encephalopathies including bovine spongiform encephalopathy,
scrapie disease in sheep, and CreutzfeldtÐJakob disease in
humans. Prion disorders (also referred to as transmissible spongiform encephalopathies (TSE)) are fatal neurodegenerative diseases that are linked to misfolding and subsequent aggregation of the normal globular protein PrP$^C$. According to the "protein only hypothesis"\cite{Prusiner98PNAS,Legname04Science} the aggregated scrapie form PrP$^{SC}$ is the causative agent of the various TSE linked diseases. The scrapie conformation can recruit the cellular form PrP$^C$ and facilitate its conversion to PrP$^{SC}$, thus ensuring self-propagation\cite{Griffith67Nature}. Given the crucial role played by the misfolded states of PrP$^C$ in TSE it is natural that there has been intense effort in deciphering the mechanism of conversion from the normal cellular form to the PrP$^{SC}$ state. 

It is believed that residues 90-231 of PrP$^C$ are the minimal infectious unit. Structures of mammalian as well as non-mammalian PrP$^C$ from a number of species show that the residues 90-121 are mostly disordered while the rest of the residues are ordered\cite{Riek96Nature,Zahn00PNAS,Calzolai05PNAS}. The structured C-terminal part of PrP$^C$ consists of three helices H1, H2, and H3  (Fig.~1) and two small $\beta$-sheets\cite{Riek96Nature,Glockshuber97TIBS}. In the mouse PrP$^C$, shown in Fig.~1, H1, H2, H3 span residues 144-153, 172-194, and 200-224, respectively.  There is no clear structural model for the scrapie form\cite{Diaz-Espinoza12NSMB} although most recent studies\cite{Cobb07PNAS,Tycko10Biochem} favor a parallel in-register arrangement of a conformationally altered form of PrP$^C$. It is known that  PrP$^{SC}$ has substantial $\beta$-strand content, which implies that in the PrP$^C$ $\rightarrow$ PrP$^{SC}$ transition a large scale conformation rearrangement must occur. 

By integrating several experimental and computational studies it has been proposed that prion aggregation is preceded by the conversion of PrP$^C$ to a monomeric aggregation-prone state PrP$^{C*}$, which unlike in the aggregation of other disease related proteins such as A$\beta$\cite{Tarus06Jacs}, is more stable than PrP${^C}$. In other words, under normal operating conditions the functional form PrP$^C$ could be metastable\cite{Thirum03COSB}. A large free energy barrier (exceeding 20-25 Kcal/mol)  separates the isoforms PrP$^C$ and PrP$^{C*}$, and hence the latter is rarely populated during the typical life cycle of PrP$^C$\cite{Baskakov01JBC}. A key question is what are the regions in PrP$^C$ that harbor residues that are most susceptible to  conformational changes in the $PrP^C \rightarrow PrP^{C*}$ transition?   Several years ago Dima and Thirumalai (DT)\cite{Dima02BJ,Dima04PNAS}  proposed that in mammalian prions the core of the ordered C-terminus region of PrP$^C$ is frustrated, and the associated instability could trigger a $\alpha \rightarrow \beta$ transition.   Frustration implies that the secondary structures adopted by certain residues in the native state are incompatible with their natural propensities as assessed by comparison to a database of structures. Using bioinformatics methods, structural analysis, and molecular dynamics simulations DT showed that conformational fluctuations in the C-terminal end of H2 and in large portion of H3 are involved in the $PrP^C \rightarrow PrP^{C*}$ transition in mammalian prions\cite{Dima02BJ,Dima04PNAS}.  Because global conformational change is required to populate the aggregation-prone PrP$^{C*}$ state the barrier to its formation is large, which explains the rarity of prion disorders during the normal function.

Although prion genes are shared by vertebrates, non-mammalian species are apparently not susceptible to prion disorders.  By studying the turtle prion protein, 
Simonic {\it et al} suggested that $\alpha$-helix$\rightarrow$$\beta$-sheet transition is unlikely
in non-mammals \cite{Simonic00FEBS}.  Using several structural measures and a quantitative assessment of frustration based on the concept  that certain sequences are discordant\cite{Kallberg01JBC} (they adopt a  certain secondary structure ($\alpha$ helix for example)  in a protein but would normally have a different structure ($\beta$ strand) in a majority of proteins) DT  showed that the avian helices are not as frustrated as their mammalian counterparts \cite{Dima02BJ}. This study and a related work\cite{Dima04PNAS} rationalized the finding that non-mammalian species typically do not acquire prion disorders.

In order to provide further insights into the extent of local frustration we use a sequence-based method to tease out the plausible reasons for the differences in mammalian and non-mammalian PrP$^C$. In particular, we applied the statistical coupling analysis (SCA) technique \cite{Suel03NSB,Hatley03PNAS,Shulman04Cell}
to extract a network of residues which are evolutionarily important from
multiple sequence alignment of the protein family. We performed SCA for prion proteins from
mammals and non-mammals  separately and then analyzed the  
networks  of covarying residues from the perspective of function.   Although structurally similar, the differences in the degree of frustration in parts of H2 and H3 results in these regions being the likely initiation sites for PrP$^C \rightarrow$ PrP$^{SC}$ transition in mammalian prions. 

The conclusions obtained from the sequence-based SCA are complemented by probing the response of mPrP$^C$  human PrP$^C$ (huPrP$^C$) to mechanical force without and with disulfide (SS) bond between Cys179 and Cys214. By generating a number of unfolding trajectories we generated a free energy profile, $G(R)$, as a function  of the molecular extension $R$. The profile and the dynamics of rupture of contacts clearly delineate  the order of unfolding. The instability associated with residues in H2 and H3 results in their unfolding prior to the more stable parts.   Although there are differences in the mechanical stability of PrP$^C$ under reducing conditions (no SS bond) and oxidizing conditions (SS bond intact) the initial unfolding, which is needed to access PrP$^{C*}$, is localized in H2 and H3.  Both the evolutionary based analysis and responses to  mechanical force show that the initial transition in the conversion from cellular form to the scrapie form must involve conformational changes in the C-terminal helices H2 and H3. The results using the SCA  also explain the absence of PrP$^{Sc}$ formation in non-mammals. 
   
\section{Methods}

\textit{Statistical Coupling Analysis:}In order to identify the network of residues that are evolutionarily
related, we use our formulation \cite{Dima02ProtSci,Chen07JMB} of the Sequence-based Statistical Coupling 
Analysis (SCA) introduced by Lockless and Ranganathan in their pioneering studies
 \cite{Suel03NSB,Hatley03PNAS,Shulman04Cell}. The SCA is remarkably versatile and provides physically meaningful results provided the data base of sequences is large\cite{Zhenxing09Proteins}. We first created a multiple sequence alignment of the PrP$^C$ sequences. A statistical free energy-like
function at each position, $i$, in a multiple sequence alignment (MSA) is defined as
\begin {equation}
\frac{\Delta G_i} {k_{B}T^*}=\sqrt{\frac{1}{C_i}\sum_{x=1}^{20}{[p_i^xln(\frac{p_i^x}{p_x})]^2}}  
\label{eq:delta_Gi}
\end {equation}
where, $k_{B}T^*$ is an arbitrary energy unit, $C_i$ is the number of types of amino acid that appears at position $i$, $p_x$ is the mean frequency of 
amino acid $x$ in the MSA. In eq. \ref{eq:delta_Gi} $p_i^x=\frac{n_i^x}{N_i}$, where $n_i^x$ is the number of times amino acid $x$ appears at position $i$ in the MSA, 
and $N_i=\sum_{x=1}^{20}n_i^x$.

The basic hypothesis of the SCA is that correlation or covariation between two positions $i$ and $j$ may be inferred by comparing the statistical properties of the MSA and a 
sub-alignment of sequences (derived from the MSA) in which a given amino acid  at position $j$ is conserved ($S_j=0$). 
The restriction that $S_j=-\sum_{x=1}^{20}p_j^xlnp_j^x = 0$ in the sub-alignment is referred to as sequence 
perturbation at position $j$\cite{Suel03NSB}. The effect of perturbation is assessed using,
\begin {equation}
\frac{\Delta \Delta G_{ij}}{kT^*}=\sqrt{\frac{1}{C_i}\sum_{x=1}^{20}{[ p_{i,j}^{x}ln(\frac{p_{i,R}^{x}}{p_x})-p_i^xln(\frac{p_i^x}{p_x})]^2}}
\label{eq:delta_delta_Gij}
\end{equation}
where $p_{i,j}^x=n_{i,j}^x/N_{i,j}$, $n_{i,j}^x$ and $N_{i,j}$ are the number of sequences in the sub-alignment in which $x$ appears in the $i^{th}$ 
position and $N_{i,j}=\sum_{x=1}^{20}{n_{i,j}^{x}}$. The coupling between sites $i$ and $j$ inferred using Eq.\ref{eq:delta_delta_Gij} differs from the original formulation, which has little consequence on the qualitative conclusions\cite{Dima06ProtSci}. Our procedure, which is a generalization of the sequence entropy, has been successfully used to identify allostery wiring diagram in enzymes\cite{Chen07JMB}. 

In order to obtain statistically meaningful results using the SCA, it is important to choose 
the sub-alignments appropriately \cite{Dima06ProtSci}. Let $f=p/N_{MSA}$ where $p$ is the number of sequence in the sub-alignment and $N_{MSA}$ is the total number of 
sequences in the MSA. We choose $f$ = 0.35 to satisfy the central limit theorem \cite{Dima06ProtSci} ensuring  that the statistical properties of the 
sub-alignments coincide with the full MSA. Using $f=0.35$, we calculated the matrix elements 
$\Delta\Delta G_{ij}$ which estimate the response of position $i$ in the MSA to all allowed perturbations at $j$ ($S_j=0$). The rows (labeled $i$) 
in $\Delta\Delta G_{ij}$ correspond 
to positions in the MSA. We determined the network of covarying residues using the elements 
$\Delta\Delta G_{ij}$ in conjunction with coupled two-way clustering algorithm \cite{Getz00PNAS}. The extent to which the rows 
$\Delta\Delta G_{ij}$ and $\Delta\Delta G_{kj}$ are similar is assessed using the Euclidean measure \cite{Dima06ProtSci}.  
Because $\Delta\Delta G_{ij}=0$ for perfectly conserved positions and for sites where the amino acids are found at their mean frequencies in the MSA 
($p_i^x=p_i$), the SCA cannot predict the role these residues might play in the function or dynamics of the enzyme.  
\\

\textit{Self-organized polymer (SOP) model for prion protein:} In order to study the instabilities in the ordered regions of PrP$^C$ we simulated the effect of mechanical force using the coarse-grained self-organized polymer (SOP) model, which has been used with considerable success in predicting the outcomes of single molecule force spectroscopy of proteins and RNA \cite{Mickler07PNAS,Hyeon07BJ,Lin08JACS} as well as in describing complex dynamical processes ranging from protein folding to allosteric transitions in proteins\cite{Hyeon06Structure,Hyeon06PNAS,Chen07JMB}.  Because force does not alter the interactions involving the protein of interest the response of proteins to force is particularly well suited to probe specific regions of instability. In the simplest version of the  SOP model the structure of a protein is represented using only by the $C_{\alpha}$
coordinates, $r_i (i=1,2,...N)$ with $N$ being the number of
amino acids. The potential energy of the prion protein in the SOP representation is 
\begin {eqnarray}
\label{eqn:potential}
H(\{r_i\})&=&V_{FENE}+V_{NB}^{ATT}+V_{NB}^{REP}\\\nonumber
       &=&-{\sum_{i=1}^{N}{
                           \frac{k}{2} R_0^2
                           \log(1-\frac{(r_{i,i+1}-r_{i,i+1}^{0})^2}
                                        {R_0^2})}} \\ \nonumber
       &+&{\sum_{i=1}^{N-3}\sum_{j=i+3}^{N}\epsilon_h[(\frac{r_{ij}^{0}}{r_{ij}})^{12}
-2(\frac{r_{ij}^{0}}{r_{ij}})^6]\Delta_{ij}}\\\nonumber
       &+&({\sum_{i=1}^{N-2}\epsilon_l(\frac{r_{i,i+2}^{0}}{r_{i,i+2}})^{6}
+\sum_{i<j}\epsilon_l(\frac{\sigma}{r_{ij}})^{6}(1-\Delta_{ij})})
\end{eqnarray}
where   the distance between two
adjacent $C_\alpha$-atoms is $r_{i,i+1}$ and
$r_{i,j}$ is the distance between the $i^{th}$ and $j^{th}$
$\alpha$-carbon atoms and $r_{i,j}^{0}$ is
the corresponding distance between the $i^{th}$ and $j^{th}$ $C_\alpha$-atom in the folded structure.
The first term in Eq. (\ref{eqn:potential}), the finite extensible
non-linear elastic (FENE) potential, accounts for chain connectivity.
The stability of the protein is described by the non-bonded
interactions (the second term in Eq.(\ref{eqn:potential})) that assigns
attractive interaction between two residues that are in contact in the native structure. Non-bonded interactions
between residues that are not in contact in the native structure are taken to be purely
repulsive (the third term in Eq. (\ref{eqn:potential})). The value of
$\Delta_{i,j}$ is 1 if $i$ and $j$ are in contact in native structure, and is zero
otherwise. A native contact implies that the distance between the
$i^{th}$ and $j^{th}$ interaction centers is less than a cut-off distance $R_C$ (0.8nm  in this study).

The spring constant, $k$, in the FENE potential (the first term in Eq.(\ref{eqn:potential})) for stretching a covalent bond is
2,000 kcal/(mol$\cdot$nm$^2$), and the value of $R_0$, which gives the allowed extension of the covalent bond, is 0.2 nm.
The values of the parameters $\epsilon_h$, $\epsilon_l$, and $\sigma$ are taken to be 1.2 kcal/mol, 1.0 kcal/mol, and .38 nm. Because there are only a few parameters in the SOP energy function we can exhaustively explore the physical processes governing the unfolding of prion protein under tension.

\textit{Simulations:}
We assume that the dynamics of the system can be described using the Langevin
equation in the overdamped limit. The equation of motion  for the $i^{th}$ $\alpha$-carbon atom is given by
\begin {equation}
\label{eqn:dynamics}
-\zeta {\frac{\partial \vec r_i}{\partial t}}=-{\frac{\partial H {r_i}}{\partial \vec r_i}}+\vec \Gamma_i(t),
\end{equation}
where $\zeta$ is the friction coefficient, and $\Gamma_i(t)$ is a random force with  white noise spectrum. We start the Brownian dynamics simulations  by first equilibrating the prion protein at $T=300K$. Subsequently, an external force is applied to the C-terminal end of the prion protein  while the N-terminal is fixed.  By symmetry the direction of pulling does not affect the calculation of scalar quantities.  In the constant loading rate simulations force is continuously increased by attaching a spring (mimicking the harmonic trap in a laser optical tweezer experiment (LOT)  or a cantilever in AFM experiments) with a spring constant $k_s=0.15$ pN/nm. Thus, the effect of applying force to the C-terminus of the protein leads to an external mechanical force $f(t) = -k_s(z_c^0 - v_s t)$ where $z_c^0$ is the initial position of the C-terminus $\alpha$-carbon atom, $v_s$ is the pulling velocity. We use $v_s=6.4\times10^3$ nm/s, $k_s=0.15$ pN/nm. Thus, the  loading rate, $r_f = k_sv_s$ = 960 pN/s, which is comparable to the range used in typical AFM experiments but is about (5-100) times larger than the  values in LOT experiments. 

We estimate the time scale involved in the unfolding of prion protein using typical values of the friction coefficient (\ref{eqn:dynamics}) and energy scale in the SOP energy function  (\ref{eqn:potential}), which yields  $\tau_H\approx\frac{\zeta_H \epsilon_h}{k_BT_s}(\tau_L)^2$. For our choice of parameters we obtain  $\tau_L=4$ ps\cite{Hyeon06PNAS, Chen07JMB,Veitshans97F&D} with $\epsilon_h=1.2kcal/mol$, $\zeta_H=100/\tau_L$, and $T_s=300K$. The integration time step of $h=0.05\tau_H$, and using the natural measure of time for the overdamped condition for $\tau_H$  gives $h =40$ ps. 

\textbf{\textit{Fraction of contacts:}}
In order to describe the order of  unfolding of various structural elements of PrP$^C$ upon application of force we calculated the  time-dependent fraction of contacts between secondary structures, which is defined as $f_{C}(t)=\frac{\sum_{i=1,L}{N^i_{C}(t)/N^{o}}}{L}$, where $i$ is the index of the trajectory, $L$ is the total number of trajectories, $N^i_{C}(t)$ and $N^{o}$ are, respectively, the number of native contacts in the $i^{th}$ trajectory at time $t$, and the number of native contacts in the crystal structure. We calculated $f_{C}(t)$ for the four groups $\beta$1-$\beta$2, $\beta$2-H2, $\beta$2-H3, and H2-H3.  In addition, we also calculated $f_C(t)$ for the three helices. 

\textbf{\textit{Energy landscape:}}
The free energy profile as a function of end-to-end distance $R$ is calculated using $G(R)=-k_BTlog(P(R))$, where $P(R)$ is the normalized probability of $R$ over a total number of $M$ ($M$ is 200 in our case) trajectories.  At a constant loading rate, the extension of the protein $R$ increases (decreases) with the increasing (decreasing) pulling force $f$. We count, for each of the $M$ trajectories, the number of conformations, $N^i(R)$, with a certain value of $R$  where $i=1...M$. We calculate  $P(R)=\frac{1}{M}\sum_{i=1}^{M}{N^{i}(R)}$ from which $G(R)$ is readily obtained. Because the calculations are performed at a constant loading rate the calculated $G(R)$  does not represent an equilibrium free energy profile.   The free energy profile together with the dynamics of loss of structure give information about the order of unfolding of the secondary structural elements and hence gives quantitative information about the regions of instability.

\section{Results}

In order to get multiple sequence alignment (MSA) for mammals and
non-mammals, we first searched the non-redundant protein sequence database
using the PSI-BLAST program. For mammals, the prion protein sequence from Mus musculus (mouse) is used
as a query sequence and only mammalian sequences are searched. In all, 454
sequences are obtained after convergence (additional rounds of iterations in PSI-BLAST yield no new sequences)  for mammals. Similarly, trionyx sinensis (Chinese soft shell turtle)
sequence is used to identify the non-mammalian sequences, and
43 sequences are saved after convergence. We manually curated the sequences to eliminate those
that are too long or too short containing large gaps in the MSA. The resulting sequences were aligned using ClustalW \cite{Chenna_NucleicAcidsRes_2003}. With this procedure, our MSA consists of 342 sequences for mammalian PrP$^C$ and only 21 sequences
for non-mammalian prions. The small number in the MSA for non-mammalian prions does add uncertainty to the analysis of non-mammalian prions.  However, given the stark differences in the SCA predictions between non-mammals and mammals we believe that the qualitative conclusions should be robust.

\textit{Residues in the signal domain correlate more with other
regions of PrP$^C$ in mammals than non-mammals:}
The clustered residues, obtained using the SCA,  are shown in Table 1 for mammalian and
non-mammalian prion proteins.  The identities and the positions of the amino
acids in the network of covarying residues  are labelled according to their positions in the mouse PrP$^C$ (Fig. 2).  In mammals, the clustered N-terminal residues, Met1, Ala2, Asn3, Leu4, Tyr6, Leu8, Met15,
Val19 are part of the endoplasmic reticulum targeting the signal peptide
which directs the post-translational transport of prion protein into the
plasma membrane \cite{Simonic00FEBS}. In contrast, there is only one residue, Lys24
located in the cleavage site that is involved in the signal domain for non-mammalian prion proteins (see Table 1). We surmise  that the signal domain in non-mammals
is not as conserved as it is in the mammalian counterpart. These differences suggest that mammalian and non-mammalian prions could have different cellular functions.

\textit{Mammalian sequences are more conserved in the redox-related region than sequences from non-mammalian PrP$^C$:}
In mammals, residues Thr94, Asn96 and Val111,  neighbors of
residues His95 or His110,  are highly correlated with the residues in the octarepeat region of the
unstructured, highly flexible N-terminus that is asserted to bind copper. In contrast, 
in non-mammals, only one residue, Lys109, is
included in the cluster (Table 1).  These differences are also reflected in the sequence differences between mammalian and non-mammalian prion proteins in this region (Fig.~2a). For example, mammalian PrP$^C$  contains a number of Gly residues where as in non-mammalian prion proteins there are fewer Gly residues. In addition, there are greater variations in this region in non-mammalian prions than in mammalian PrP$^C$.  It is unclear if the highly conserved region, which is structurally disordered, plays any significant role in the conversion process. The reduced flexibility in non-mammalian PrP$^C$  tidily explains the observation that the structure of
the N-terminal region of non-mammalian PrP$^C$ is stable,
protease-resistant, and does not bind copper \cite{Simonic00FEBS}. The differing behavior of
mammalian and non-mammalian prion proteins, with regard to copper-binding related
redox reaction, supports the hypothesis that copper binding may not be the
primary function of prion protein \cite{Marcotte_Biochem_1999}.   We cannot rule out the possibility that the
emergence of prion disease is related to the metal-induced
redox reaction \cite{Barnham_TIBS_2006,Ji_TIBS_2007}, which has been argued to be a common mechanism in initiating both Alzheimer's disease and prion disorders. 

It has been suggested that the presence of the transmembrane binding motifs GxxxG (Fig.~2) in mammalian prions in the region 110-130 covering M128 is essential in triggering prion disorders.  Here, we find that the GxxxG motifs are also present in non-mammalian prions (see Fig.~2) in the same region. In mammalian prions M128 is highly conserved whereas it is less so in non-mammlian prions (Fig.~2). The similarities in the properties of the sequences in this region between mammals and non-mammals suggest that this region may not  encode for the initiating sites in the PrP$^C \rightarrow$ PrP* transition.   However, it is well known that a common polymorphism at this position in huPrP$^C$ has strong influence on the kinetics of fibril formation \cite{Lewis06JGenVirol} even though there is very little manifestation of this behavior at the monomer level. For example, polymorphism does not alter the efficiency  of conversion from the cellular form to PrP*. The differences are evident only in the formation of the critical nucleus and beyond \cite{Lewis06JGenVirol}.  Thus, although polymorphism may not influence the earliest transitions clearly they affect the kinetics of fibril formation, which is beyond the scope of the present study.  

In addition to the difference described above,  Table 1 also shows that residues Lys100, Lys103, Thr106, Asn107, Lys109, Ala112, Ala115, Ala116, Ala117, Ala119, Val120, and Val121 are not clustered in
mammals. Indeed, these residues are highly conserved in mammals and do not covary with
 other regions of the prion proteins (Fig.~2a). On the other hand, in non-mammals, these
residues are evolutionarily related (Table 1).   Finally, residues in H2 and H3 are conserved to a greater extent in mammals than in non-mammals (Fig.~2b). In addition, the stretch of TTTT  in H2 is rare, and is highly conserved in mammalian prions. This pattern of T residues has a great propensity to be in a $\beta$-strand conformation in a majority of proteins \cite{Dima02BJ}. However, they are part of a helix in mammalian PrP$^C$, thus making it unusual. Taken together, these results suggest that several residues in H2 and H3 are frustrated in a helical state, and hence are likely to be part of the initiating sites in the PrP$C \rightarrow$ PrP* transition.

\textit{Clustered residues in the C-terminus are delocalized in mammals but form stable localized
  interactions in non-mammals:}
The NMR structure in Fig~1 (PDB entry: 1AG2) shows that the C-terminal of $PrP^C$  has
three $\alpha$-helices and a two-stranded anti-parallel
$\beta$-sheet\cite{Riek96Nature,Donne_PNAS_1997}.  Covarying residues in the
network of mammals and non-mammals are shown with spheres on the same structure
for comparison (green for residues clustered in mammals and red  for
non-mammals). Clearly, all the
red spheres are localized near  the center of mass of the protein. The residues in green
are distributed in the peripheral region (the cartoon
representation of the protein chain is colored according to their 
distance to the center of mass with red
being the closest and the blue being the farthest). It appears that
the residues in the center of non-mammalian prion protein are
evolutionarily-related in order to maintain a stable
structure.  We conclude that these residues are not frustrated and the corresponding sequences are concordant implying that the $\alpha$-helical secondary structures adopted by these residues are compatible with the theoretically predicted structures for these sequences. 

Our previous studies predicted that as a result of instabilities in the dynamics  of the helical
fragments localized in the second half of H2 and parts of H3 they would undergo a transition
from $\alpha$ helical conformation to a $\beta$ and/or random coil
state \cite{ Dima02BJ,Dima04PNAS} during the PrP$^C$ to PrP$^{Sc}$ transition. The current work shows that the clustered residues in non-mammalian sequences are   located in the stable
helical fragment (Asn152, Arg155, Val175, Asn180, Val208). In contrast, for
mammals, all the clustered helical residues are part of the frustrated helices H2 and H3. For example,  Ile183 is close to the second half of H2, and residues Val202, Met204, Glu218, Lys219 and
Asp226 are part of H3. These results confirm the earlier predictions that the frustrated regions localized in H2 and H3 are most  susceptible  to conformational change, and could be designated as initiation sites in the PrP$^C \rightarrow$ PrP$^*$ transition. In contrast, we predict that H2 and H3 in PrP$^C$ from trionyx sinensis, a non-mammalian species, are not as frustrated thus explaining the lack of PrP$^{Sc}$ formation in these species.

\textit{Forced-unfolding of the mPrP$^C$ and huPrP$^C$ starts from H3 and H2.}
 In order to complement the 
 predictions based on evolutionary imprints using the SCA   we also carried out Brownian dynamics simulations (see Methods) to unfold mPrP$^C$ and huPrP$^C$ using mechanical force. We first describe results for prion proteins without the disulfide bond. At a constant loading,  mPrP$^C$ unfolds in two distinct steps (black trajectory in Fig.~3a). When $f\approx$ 35 pN, the molecular extension of the prion protein, $R$, increases by $\sim$ 10 nm. This step is associated with the rupture of H3 and H2. In the second step, at $f\approx$ 40 pN, $R$ increases from 15 to 20 nm and is associated with unfolding of the two $\beta$-strands (Fig.~1) and H1. 
 Using the force-extension curves from about 100 unfolding trajectories we calculated  $G(R)=-k_{B}T$ln$P(R)$ where $P(R)$ is the distribution of  $R$. The free energy profile $G(R)$ (Fig.~3b) shows that there are two major steps in the unfolding of PrP$^C$. When chain extension exceeds the distance between the folded state ($R$=2.3 nm) to the first barrier that is $\sim$ 5 nm (Fig.~3b) away,  H3 and H2 unfold. By  extrapolating the estimated barrier to unfolding obtained at roughly $f \sim$ 35pN to zero force using $G(R|f = 35 pN) = G(R|f=0) - \Delta R f)$ where $\Delta R$ = (5 - 2.3) = 2.7 nm we obtain that the barrier at $f$=0 would be $\approx$ 19kcal/mol, which is remarkably similar to that estimated in experiments \cite{Baskakov01JBC}. The minimum at $R \sim$ 13 nm corresponds to an intermediate state, which corresponds to conformations with H2 and H3 unfolded. The total number of residues in H2 and H3 is 48, which implies that at full extension the length gain due to their unfolding should result in $R \approx$ 18 nm assuming an extension of $a \approx$ 0.38 nm per amino acid. However, we find that that upon rupture of H2 and H3 at $f\approx$ 35 pN the gain in length is $R \approx$ 12 nm, which implies that there is residual helical structure upon stretching these segments (see the conformations in Fig.~3).  Incomplete stretching has also been reported in other helical proteins\cite{Gebhardt10PNAS}.   The second barrier at $R \sim$ 18 nm represents extension involving H1 and rupture of contacts between the two $\beta$-sheets in the N-terminal of the prion structure.
 
 Force-induced unfolding results for huPrP$^C$ (Fig.~4a) obtained using the structure (PDB code 1QLX) are shown in Fig.~4. The length of H3 in huPrP$^C$ is longer than in mPrP$^C$, which results in $R$ of the native state being longer than in mPrP$^C$. Taking this fact into account we find that the calculated $G(R)$ profiles (compare Fig.~3b and Fig.~4b) are similar. Just as in mPrP$^C$, when $R$ exceeds the first barrier located $\sim$ 4.5 nm away from the folded state, H2 and H3 unfold, and populates an intermediate state.  Both the profiles clearly show that unfolding occurs through an intermediate, at $R \sim$ 10 nm from the folded state, in which interactions involving H2 and H3 are disrupted just as in mPrP$^C$. It is also interesting to note that sequence effects are manifested in the finer details of $G(R)$ indicating that single molecule pulling experiments can be profitably used to tease out the differences between various prion proteins.  Thus, the free energy profiles, including the barrier height separating the folded and the intermediate states, are similar. These results are not surprising given that the structures of mPrP$^C$ and huPrP$^C$ are homologous.

\textit{Dynamics of force-induced loss of tertiary interactions:} The contact map for the ordered C-terminal portion of mPrP$^C$ shows (Fig.~5a) interactions between $\beta_1$ and $\beta_2$ as well those involving H2, H3, and $\beta_1$ and $\beta_2$. To assess the temporal loss of these contacts upon stretching we calculated the time-dependent decrease in the  fraction of contacts during the unfolding process (see Methods).  At $t=0$, the fraction of contacts $f_{C}(0)$  involving $\beta$2-H3 is $\approx$0.5, meaning that almost half of the native contacts involving these elements are absent at room temperature.  Similarly, for $\beta$2-H2 and H2-H3, $f_{C}(0)$ is $\approx$0.6.  The equilibrium value of $f_{C}(0)$ involving $\beta$1-$\beta$2 is $\approx$ 0.8.  The time-dependent decrease in $f_{C}(t)$ involving these secondary structural elements upon application of  force is shown in Fig~5b.  We find that the loss of contacts between H3 and the $\beta$2 occurs first (Fig~5b), followed by the rupture of the contacts between H3 and H2, and H2 and $\beta$2.  Interactions between $\beta$1 and $\beta$2 on the N-terminal of the prion protein are the most stable, and are only disrupted during the last stages of unfolding. Interestingly, the helical structure of H1 is relatively intact even after complete disruption of structure in the rest of the molecule.  If the entire C-terminal region of PrP$^C$ with the number of amino acids, $N$ = 111, is fully extended we expect $R \approx (N-1)a \approx$ 42 nm.  However, we find that even at $f \approx$ 75 pN, $R$ falls short of 42 nm. Near full extension, realized only upon stretching of H1, occurs when $f \approx$ 120 pN. This shows that H1, stabilized by salt-bridges, is unlikely to undergo conformational changes in the early stages of the PrP$^C \rightarrow$ PrP$^{Sc}$ transition.  

 The dynamics of rupture of tertiary contacts in huPrP$^C$ (Fig.~4c) is nearly quantitatively identical to that observed in mPrP$^C$. Here, interactions involving H3 and H2 are disrupted prior to the rupture of contacts of $\beta_1$, $\beta_2$, and finally H1. Thus, based on pulling simulations of mPrP$^C$ and huPrP$^C$ we conclude that the major instabilities are localized in H2 and H3.

 \textit{Forced unfolding of mPrP$^C$ with intact SS bond:} Mammalian prions contain an internal disulfide bond between Cys179 and Cys214 that tethers H2 and H3 to each other (Fig.~6a), thus enhancing the stability of the region around the SS bond. We carried out Brownian dynamics simulations to assess the influence of $f$ on the internal stability of mPrP$^C$ with SS bond present. In these simulations the covalent the SS bond is modeled by adding a stiff FENE potential (first term in Eq. (3)) between Cys179 and Cys214 with $k$ =2,000 kcal/mol$\cdot$nm$^2$. 

The free energy profile $G)R)$ in Fig.~6b shows that with intact SS bond the entire PrP structure is more stable, and the intermediate state at $R \sim$13 nm found in mPrP$^C$ is absent (Compare Figs.~2b and 5b)). Due to the SS restraint, the helical contents of H2 and H3 between Cys179 and Cys214 remain intact throughout  the simulations. However, the helical structures outside the region surrounding the SS are less stable, and are the first to rupture. As shown in Fig.~6c, in the early stage of pulling simulations (t=0 to 20 ms), the fraction of contacts in the first half of H2 (residue 171-179 denoted by H2*) decreases. Residues in the second half of H3 (residue 214-223, labeled H3*) lose a large fraction of their contacts. In contrast, the fraction of contacts in H1 remains to  0.7.  Thus, there is a consistency in the extent of frustration in regions associated with H2 and H3 both with and without SS bond.

\section{Discussion} 

Although the structures of a number of species of PrP$^C$ have been determined the sequence of events that drive the monomer to scrapie form is not well understood.   From both sequence and structural analyses\cite{Dima02BJ}, experiments\cite{Lu07PNAS,Kuwata02Biochem},  molecular dynamics simulations\cite{Dima04PNAS}, and the response to mechanical force (Figs.~3 - 6) it is clear that  H1 is stable. In mammalian prions the stability arises because of perfect placement of oppositely charged residues at locations $i$ and $(i+4)$\cite{Dima04PNAS}. Such an arrangement is rarely, if ever, found in proteins in the genomes of in {\it E. Coli} and yeast genomes\cite{Dima04PNAS}.  More importantly,  experiments using CD and NMR \cite{Liu99Biopolymers,Ziegler03JBC} show that the isolated H1 is extremely stable with high degree of helix content over a wide range of solvent conditions. Using helical constructs from mPrP$^C$, with a few flanking residues that apparently do not have any influence on the helix population of the interior residues, it was demonstrated \cite{Liu99Biopolymers} that H1 has high intrinsic helix propensity. In a later study \cite{Ziegler03JBC} probed the stability of isolated H1 from huPrP$^C$ over a broad range of solution conditions. Surprisingly,  the intrinsic helix content is nearly 60\%, which is unusual given that there are no long range tertiary interactions to stabilize the isolated H1. Both these studies \cite{Liu99Biopolymers,Ziegler03JBC} assert that H1 is unlikely to be involved the conversion process to the scrapie form with the latter \cite{Ziegler03JBC} emphasizing  that the stability of H1 could be a barrier in the PrP$^C \rightarrow$ PrP* transition. These observations the isolated  suggest  that, at least in the early stages, it is unlikely H1 would undergo conformational changes.  It should be noted that others have proposed a key role for H1 in initiating the PrP$^C$ to PrP$^{Sc}$ conversion\cite{Morrissey99PNAS,DeSimone07BJ}.   Our findings and several experiments (see below) strongly suggest that the  conformational changes in the stable H1 is not the dominant feature in the creation of the aggregation prone PrP* from PrP$^C$.  This conclusion does not imply that H1 does not undergo a change in conformation at later stages. However, such a possibility has been ruled out in certain recent experiments \cite{Tycko10Biochem,Lu07PNAS}.

\textit{Experimental evidence from monomer dynamics:} The finding that the initiation sites that drive the PrP$^C \rightarrow$ PrP* transition must involve H2 and H3 helices finds considerable experimental support.     Several experiments, probing the dynamics of mammalian PrP$^C$ and their mutants, under a variety of conditions have established that H2 and H3 undergo substantially larger fluctuations than the rest of  the structure, and thus point to their potential instability\cite{Hosszu10Biochem,Bae09Biochem,Kuwata07PNAS}. (1)Perhaps, the earliest evidence for the potential role  H2 and H3 play in creating PrP* in Syrian hamster comes from the $^15$N-$^1$H two dimensional NMR experiments \cite{Kuwata02Biochem}, which showed that in a small population of the aggregation species H2 and H3 are locally disordered. They further suggest that the transition to the PrP* state, with disordered H2 and H3, may be the key step in the association with the scrapie form. (2) More recently, Bae {\it et. al.} \cite{Bae09Biochem} have used NMR to characterize the intrinsic flexibility of mPrP$^C$ and few key mutants. By measuring the NMR order parameters they surmise that regions of H2 and H3 have smaller values of the order parameter, and hence more flexibility (see the discussion related to Fig.~4 in \cite{Bae09Biochem}). Based on this study they assert that segments that span H2 and H3 may constitute the initiating sites for pathogenic mutants as well as the wild-type. It is worth noting that amino acid sequences in the C-terminus mammalian prions are well conserved \cite{Billeter97PNAS}, which implies that the initiation sites for PrP$^C \rightarrow$PrP* transition is likely to be similar in all mammalian species.  (3) It has been argued that $\beta$-PrP$C$, an intermediate lacking SS bond created under acidic conditions, has enhanced $\beta$-strand content \cite{Hosszu09JBC}. In this conformation H2 is apparently  unstable where as there is helical content in H3. This study is not inconsistent with our conclusions. As already noted in  \cite{Hosszu09JBC}, the presence of structure in H3 given that H2 is unstable is puzzling especially considering that $\beta$-PrP$^C$ has far greater $\alpha$-helical content than the conformations adopted in the fibrils. Furthermore, the monomers in the fibril have parallel in-register $\beta$-sheet arrangement involving both H2 and H3 (see below). Nevertheless, the stability of H1 and the instability of H2 in $\beta$-PrP accord well with our findings. (4) Finally, analyses of dynamics of structural domains based on short molecular dynamics simulations \cite{Blinov09Biochem} it has been argued that H3 is unstable, which accords well with our study. However, they also suggest that H1 is dynamically unstable, which is not supported by the present study nor by experiments showing that even the isolated H1 is stable \cite{Liu99Biopolymers,Ziegler03JBC}.  More recently, Santo {\it et. al.} \cite{Santo11Prion} have shown using computations and NMR experiments in a number of mammalian prions that the largest dynamical domain is localized in H2 and H3. In addition, the dynamics associated with this region is coupled to $\beta_2$ just as found in the present study (see Figs.~3c and 4c). It is gratifying that a number of different approaches yield a consistent picture for the role of H2 and H3 in the initial stages of PrP$C \rightarrow$ PrP* transition.  As shown here fluctuations in this region arise due to decreased stability, which in turn can be traced to the unusual sequence composition in H2\cite{Dima02BJ}. Not coincidentally many of the naturally occurring pathogenic mutations are also found here. 

 Destabilization of  H2 and H3, which form substantial core of PrP$^C$, would result in could result in unfolding of the whole protein. As a result of near global unfolding most of the prion protein would be exposed to the solvent. Results from two NMR experiments could be used to infer that all three helices have similar stabilities based on their dynamical behavior.  (a) Equilibrium H/D exchange experiments on huPrP$^C$ done sometime ago \cite{Hosszu99NSB}   found that the protection factor for the core of the protein was essentially the same as the equilibrium constant between the folded and unfold states. However, from these equilibrium experiments the order of unfolding in individual molecules cannot be deduced nor can the population (estimated to be $\sim$ 1\%) \cite{Kuwata02Biochem} of PrP* molecules  be inferred. The initial disruption of structures associated with H2 and H3 ensures that interactions associated with H1 are destabilized rapidly, thus explaining the observed pattern of protein factors \cite{Hosszu99NSB}. (b) It has been suggested \cite{Sullivan08ProtSci} that for truncated mPrP$^C$ (residues 113-231) that all three helices have similar flexibility. Although the results of this study is not in agreement with the conclusions reached elsewhere \cite{Bae09Biochem} even these authors implicate regions in H2 as potential initiation sites. 

 \textit{Consistency with proposed fibril structures:}  Recent experiments provide convincing evidence that in the fibril state H2 and H3 have altered conformations, and  adopt $\beta$-strand structures. (1) Using H/D exchange experiments of PrP$^{SC}$ formed from huPrP$^C$ it was established that the highest protection factors were found in residues starting around 169 and encompassing H2  and H3 \cite{Lu07PNAS}. They attributed the large values of the protection factor to extended hydrogen bonded cross $\beta$ structure. In a subsequent study Surewicz and coworkers \cite{Cobb07PNAS} used site directed labeling and EPR to demonstrate that in the fibril state the core of the protein (including H2 and H3) form a single layer structures that are stacked in an  in-register parallel manner. (2) More recently, constraints obtained from solid state NMR experiments on Syrian hamster provided compelling evidence that the fibril core contains is formed from residues 173-224, which includes H2 and H3. These segments form $\beta$ strands. These experiments  and the high $\beta$-strand content in PRP$^{Sc}$ cannot be explained without invoking a critical role for H2 and H3 in the conversion process. We conclude that our results are consistent with a substantial number of experiments on  both monomers and fibrils.

\section{Conclusions} 

Our findings and experiments cited above show that the frustrated helices H2 and H3 must undergo a transition to an assembly competent state, PrP*, by adopting an extended strand conformation. It should be emphasized that we are referring to instabilities associated with H2 and H3, which cannot be inferred from equilibrium titration of PrP$^C$ in the presence of denaturants. Because such a transition involves near global unfolding of a substantial part of the protein (resulting in similar protection factors in the ordered regions of the equilibrium H/D exchange experiments \cite{Hosszu99NSB} the barrier separating PrP$^C$ and PrP* must be large\cite{Thirum03COSB} so that under normal conditions the population of PrP* is likely to be low. This proposal is consistent with the finding that even at high pressures only $\sim$ 1\% of the protein is in the PrP* state\cite{Kuwata02Biochem}. In addition, as  PrP* molecules associate and grow the strands resulting from $\alpha \rightarrow \beta$ transition in H2 and H3 would form the core of the fibril as shown in a number of recent studies\cite{Lu07PNAS,Cobb07PNAS,Tycko10Biochem,Biljan11JMB}. The resulting model, which favors formation of parallel $\beta$-strand fibrils involving conformationally altered H2 and H3 in the core, explains a number of biophysical experiments including the observation of high protection factors in the H/D exchange experiments  in the core of the fibril \cite{Lu07PNAS}. Thus, despite  the suggestion that PrP$^{Sc}$  could be described using $\beta$-helix\cite{Govaerts04PNAS} or $\beta$-spiral models\cite{DeMarco04PNAS} in which the C-terminal structures are intact (do not undergo conformational changes during the the transition to the PrP$^{Sc}$ form) majority of the recent experiments suggest a major initiation role for H2 and H3, as suggested here.  We should emphasize however that a structural model of PrP$^{Sc}$ will be needed to establish the conformational changes in PrP$^C$ that drive the cellular form to the pathogenic scrapie state.

\section {Acknowledgment:} We thank the National Science Foundation (CHE 09-4033) and the National Institutes of Health (GM089685) for supporting this work.
\newpage

\newpage

\newpage

\setcounter{table}{0}
\renewcommand{\thetable}{\arabic{table}}
\begin{table}[h!b!p!]
\caption{Networks of residues in mammalian and non-mammalian prion proteins}
\begin{tabular}{lll}
\hline
&&Residue indices as in mouse prion protein\\
\hline

Mammals &&1, 2, 3, 4, 6, 8, 15, 19, 94, 96, 111$^{a}$
\\&&{137, 165, 183, 202, 204, 218, 219, 226,227, 231, 233, 234}$^{b}$\\
\hline
Non-mammals &&24, 29, 37, 49, 100, 103, 106,,107, 109, 112, 115, 116, 117, 119, 120,
121 $^{a}$
\\&&134, 152, 155, 158, 160, 175, 180, 208, 217$^{b}$\\
\hline
&&a: Disordered NMR structure. b: Ordered NMR structure.\\
\hline
\end{tabular}
\end{table}

\newpage

\begin{center}
\textbf{\large{Figure Captions}}
\end{center}

\hspace{-1.2em}Fig.~1  Ribbon diagram of mouse prion (PDB code 1AG2). We only show the structured C-terminal region.  The spheres represent the network of covarying residues calculated using the sequence-based Statistical Coupling Analysis.  Green (red) corresponds to mammals (non-mammals). 

\bigskip

\hspace{-1.2em}Fig.~2 Alignment of  sequences for prion proteins from mammals and non-mammals. Numbering of residues corresponds to mPrP$^C$.  The sequence of mPrP$^C$ is listed at the bottom of non-mammals. To display the alignment clearly we split the sequence into two halves. (a) We show alignment for residues 1-120. (b) Residues 121-231 are shown.

\bigskip

\hspace{-1.2em}Fig.~3 (a) Force-extension curves for two trajectories generated by pulling mPrP$^C$ from the C-terminus while keeping the N-terminus fixed. The structures that unravel at various stages as force is increased are shown for the black trajectory. (b) Free energy-like profile generated using the histogram of extensions sampled in 100 pulling simulations. Representative conformations in the basin at $R \sim$ 14 nm, $R \sim$ 23 nm, and $R \sim$ 33 nm are shown.

\bigskip

\hspace{-1.2em}Fig.~4 (a) Cartoon representation of the human prion protein (PDB code 1QLX) displaying only the structured  C-terminal region. The secondary structural elements are labeled. (b) Free energy profile, $G(R$,  generated using the histogram of extensions sampled in 100 pulling simulations. Representative conformations in the basin at $R {\sim}$ 5 nm,  $R{\sim}$ 13 nm and $R {\sim}$ 25 nm are shown. (c) Time-dependent changes in the loss of fraction of contacts between different secondary structural elements labelled in the figure. 

\bigskip

\hspace{-1.2em}Fig.~5 (a) Contact map of mPrP$^C$ corresponding to the structure shown in Figure 1. Two residues are in contact if the distance between them is less than 0.8 nm. The contact map shows that H1 is peripherally located and does not form interactions with the rest of the ordered C-terminal residues. (b) Time-dependent changes in the loss of fraction of contacts between different secondary structural elements labelled in the figure. Remarkably, H1 resists mechanical force the most and is disrupted only after loss of all the interactions in the rest of the protein. 

\bigskip

\hspace{-1.2em}Fig.~6 (a) Structured C-terminal  of the mouse prion with disulfide bond shown as black dashed line.  (b) The dependence of the free energy profile generated using the histogram of extensions sampled in 100 pulling simulations as a function of the molecular extension, $R$. Representative conformations in the basin at $R {\sim}$ 2.5 nm,  R ${\sim}$ 13 nm and $R{\sim}$ 23 nm are shown. (c) Time-dependent changes in the loss of fraction of contacts between different secondary structural elements labelled in the figure. $f_{contact}$ for H1, H2$^*$ and H3$^*$ (see text for definition) are shown in the inset.

\newpage
 
\begin{figure}[ht]
\includegraphics[width=8.00in]{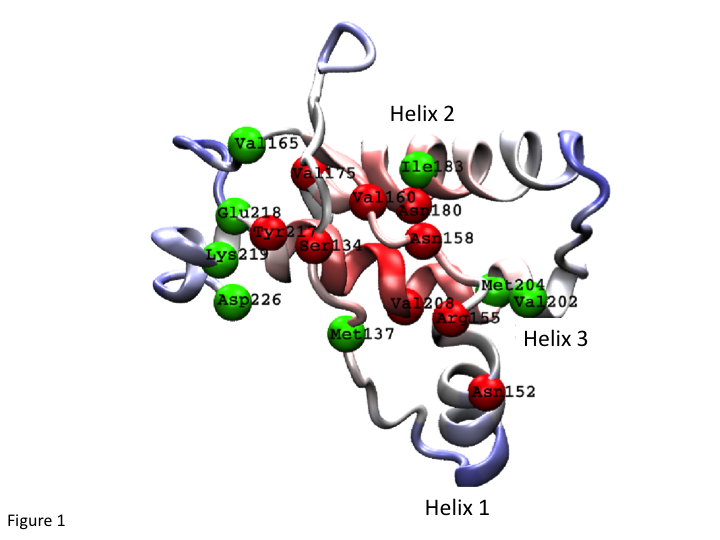}
\end{figure}

\newpage
\begin{figure}[ht]
\includegraphics[width=6.00in]{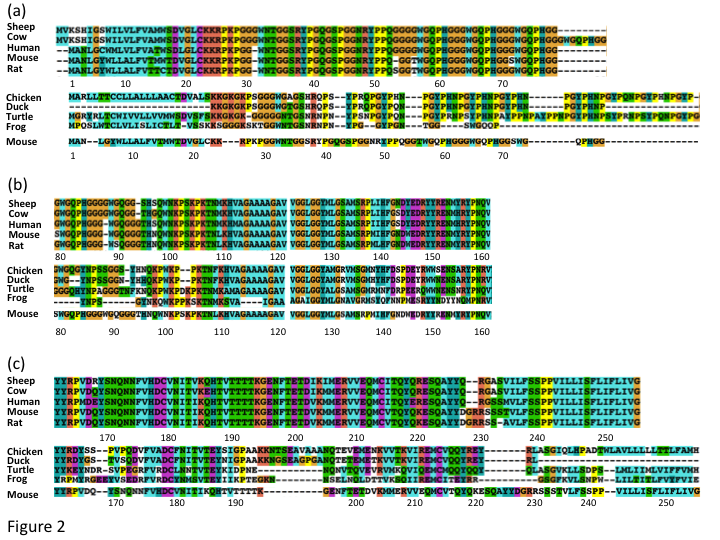}
\end{figure}

\newpage
\begin{figure}[ht]
\includegraphics[width=6.00in]{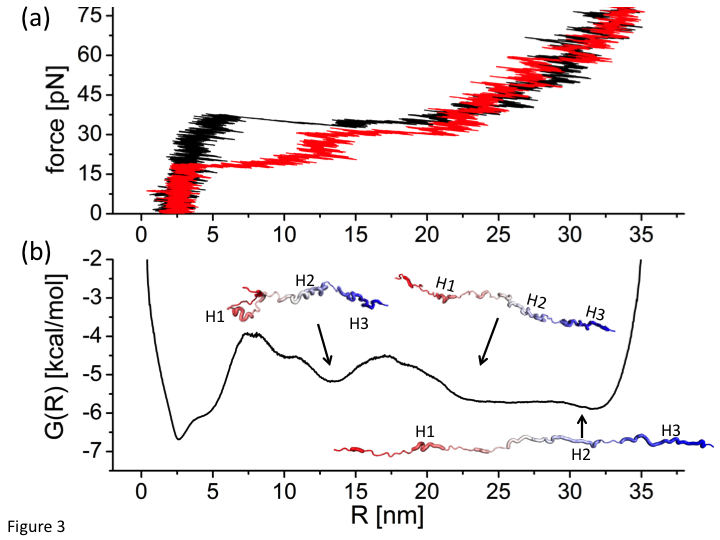}
\end{figure}

\newpage
\begin{figure}[ht]
\includegraphics[width=6.00in]{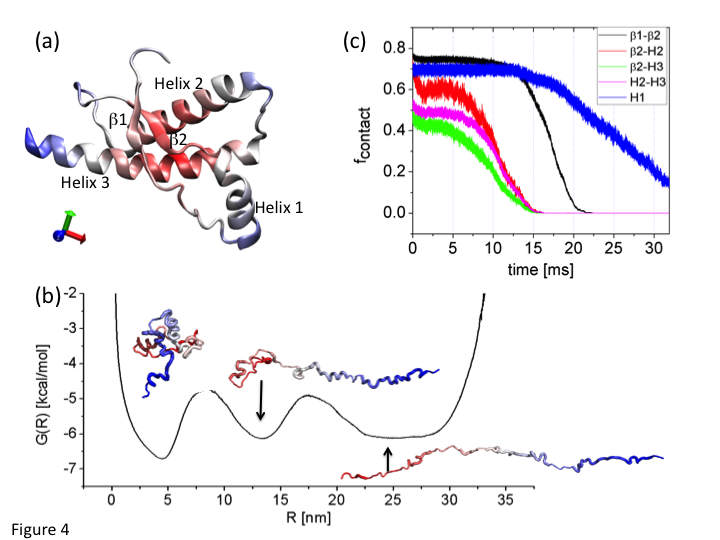}
\end{figure}
\newpage
\begin{figure}[ht]
\includegraphics[width=6.00in]{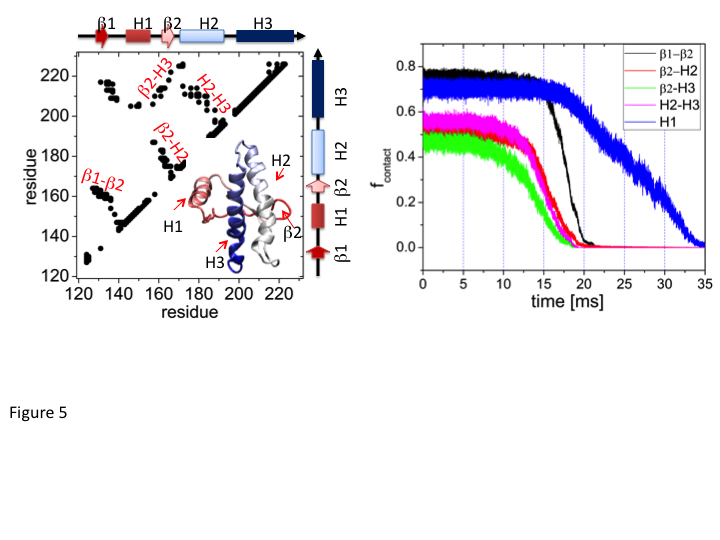}
\end{figure}
\newpage
\begin{figure}[ht]
\includegraphics[width=6.00in]{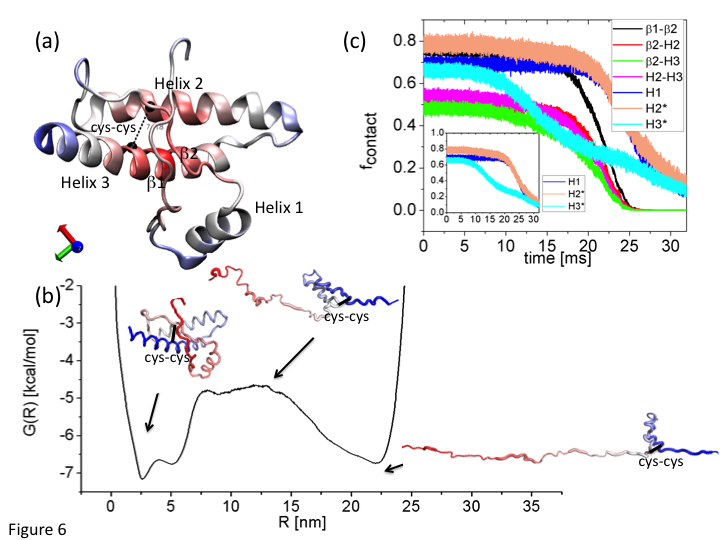}
\end{figure}

\end{document}